\documentclass[12pt]{amsart}

\usepackage{amsmath,amsfonts,amsbsy,amsthm}
\usepackage{a4wide} 

\usepackage{dsfont}
\usepackage{mathtools}
\usepackage{mathrsfs}

\usepackage{enumerate}
\usepackage{enumitem}
\setlist{leftmargin=2em}

\numberwithin{equation}{section}

\makeatletter
\newtheoremstyle{corsivo}
   {\medskipamount}{\medskipamount}%
   {\itshape}{}%
   {\bfseries}{}%
   { }
   {\thmname{#1}\thmnumber{\@ifnotempty{#1}{ }\@upn{#2}}%
    \thmnote{ {\bfseries\boldmath(#3)}}.}%
\makeatother

\theoremstyle{corsivo}
\newtheorem{theorem}{Theorem}[section]

\newtheorem{corollary}[theorem]{Corollary}
\newtheorem{proposition}[theorem]{Proposition}

\makeatletter
\newtheoremstyle{dritto}
   {\medskipamount}{\medskipamount}%
   {\rmfamily}{}%
   {\bfseries}{}%
   { }
   {\thmname{#1}\thmnumber{\@ifnotempty{#1}{ }\@upn{#2}}%
    \thmnote{ {\bfseries\boldmath(#3)}}.}%
\makeatother

\theoremstyle{dritto}
\newtheorem{definition}[theorem]{Definition}
\newtheorem{remark}[theorem]{Remark}

\newcommand{\sub}[1]{_{\mathrm{#1}}}
\newcommand{\su}[1]{^{\mathrm{#1}}}

\newcommand{\eps}{\varepsilon} 

\newcommand{\Id}{\mathds{1}}  

\newcommand{\eu}{\mathrm{e}}
\newcommand{\iu}{\mathrm{i}}   
\newcommand{\di}{\mathrm{d}}

\newcommand{\N}{\mathbb{N}}

\newcommand{\R}{\mathbb{R}}
\newcommand{\C}{\mathbb{C}}

\newcommand{\Or}{\mathcal{O}}

\newcommand{\cS}{\mathcal{S}}

\newcommand{\norm}[1]{\left\| #1 \right\|}

\newcommand{\set}[1]{ \left\{  #1 \right\}} 

\DeclareMathOperator{\Ran}{Ran}

\DeclareMathOperator{\dist}{dist}

\newcommand{\periodic}{MP}

\usepackage{etoolbox}

\makeatletter
\patchcmd{\@setaddresses}{\indent}{\noindent}{}{}
\patchcmd{\@setaddresses}{\indent}{\noindent}{}{}
\patchcmd{\@setaddresses}{\indent}{\noindent}{}{}
\patchcmd{\@setaddresses}{\indent}{\noindent}{}{}
\makeatother

\title[Vanishing of power-law corrections to Kubo's formula]{Vanishing of power-law corrections\\ to Kubo's formula for the Hall current\\ at incommensurate magnetic fields}

\author{Gabriele Mazzini}
\address[G.~Mazzini]{Gran Sasso Science Institute -- GSSI.\newline Viale Francesco Crispi 7, 67100 L'Aquila (AQ), Italy}
\email{\href{mailto:gabriele.mazzini@gssi.it}{gabriele.mazzini@gssi.it}}

\author{Domenico Monaco}
\address[D.~Monaco]{Dipartimento di Matematica ``G.~Castelnuovo'', Sapienza Università di Roma.\newline Piazzale Aldo Moro 5, 00185 Roma (RM), Italy}
\email{\href{mailto:domenico.monaco@uniroma1.it}{domenico.monaco@uniroma1.it}}
\date{January 29, 2026. arXiv v1}

\usepackage{hyperref}

\begin{document}

\begin{abstract}
We consider a non-interacting electron gas confined to a two-dimensional crystal by the action of a perpendicular magnetic field; in the one-particle approximation, the dynamics of the system is modelled by a spectrally gapped Bloch--Landau Hamiltonian. No commensurability condition is assumed between the magnetic flux per unit cell and the quantum of magnetic flux. We construct a non-equilibrium almost-stationary state (NEASS) which ``dresses'' the equilibrium Fermi projection on states below the spectral gap, and models the state of the system after the addition of a weak external electric field of strength $\eps \ll 1$. Having in mind applications to the integer quantum Hall effect, we probe the response of a current operator in the direction transverse to that of the applied electric field, and show that the resulting current density in the NEASS is linear in~$\eps$, with no power-law corrections. The linear response coefficient, namely the Hall conductivity, is computed in terms of the equilibrium Fermi projection via the double-commutator formula, in accordance with the prediction from Kubo's linear response theory.

Our results generalize the methods and findings of [G.~Marcelli, D.~Monaco, {\it Lett. Math. Phys.\ }{\bf 112} (2022), 91] to the setting of uniform magnetic fields with incommensurate magnetic flux per unit cell, and to lattice-periodic perturbation of such magnetic fields.

\medskip

\noindent \textsc{AMS MSC2020 classes:} 81Q15, 81Q20, 81V70
\end{abstract}

\maketitle

\section{Introduction}

Methods from adiabatic theory have proved successful in the endeavor to compute linear responses in quantum systems out of equilibrium, and many results have recently been obtained to set on a mathematically rigorous footing the program proposed in statistical mechanics by Kubo \cite{kubo1957statistical, henheik2021justifying}. Of particular interest, also in view of possible application in quantum technologies, is the case of the (integer) quantum Hall effect, in which the equilibrium system is a two-dimensional electron gas subject to a perpendicular magnetic field of strength~$b$, the perturbation driving it out of equilibrium is an external electric field, and the observable whose response is measured is a current in the direction transverse to the external driving. The response coefficient, to first order, is the ratio between the out-of-equilibrium current density and the driving electric field strength, namely the transverse conductivity $\sigma\sub{Hall}$, which can be computed within Kubo's setting. Contrary to the classical prediction by Hall, from which one would expect a linear dependence between~$\sigma\sub{Hall}$ and~$b$ \cite{hall1879new}, the experiments conducted by von Klitzing's group discovered that, at the quantum level (that is, at approximately zero temperature), the linear response coefficient display quantized plateaus in its $b$-dependence, occurring at integer multiples of the physical constant $e^2/h$ (thereby named after von Klitzing himself) \cite{von1986quantized}. This peculiar phenomenon paved the way for the investigation of topological effects in quantum transport, and still provides stimulating challenges for theoreticians.

This paper is concerned with a different, but related, aspect of the response of the Hall current density $j\sub{Hall} \equiv j\sub{Hall}(\eps)$, namely the fact that, while $\eps \, \sigma\sub{Hall}$ should by definition provide only a \emph{linear} approximation to the $\eps$-dependence of $j\sub{Hall}$, it turns out that no further power-law corrections are present: in formul\ae,
\[ j\sub{Hall}(\eps) = \eps \, \sigma\sub{Hall} + \Or(\eps^\infty)\,, \]
meaning that the error term vanishes faster than any power of $\eps$. This insightful observation was first made by Klein and Seiler \cite{klein1990power}, who proved the analogous statement for the Hall conductance, namely the ratio between the Hall current (rather than current density) and the external voltage drop (rather than electric field strength); this result was extended in \cite{bachmann2021exactness} to the setting of interacting fermions on a lattice. Both of these previous works use time-adiabatic pertubation theory and model the perturbing electric field by means of a slow time-dependent magnetic flux insertion {\it à-la} Laughlin. More recently, by exploiting instead space-adiabatic perturbation theory \cite{teufel2003adiabatic, teufel_non-equilibrium_2020}, the above statement on the Hall conductivity was proved both within the setting of lattice-translation-invariant one-body Hamiltonians in the continuum \cite{marcelli2022purely} and for interacting fermions on a lattice \cite{wesle2025near, marcelli2025gapped}; we refer the reader to \cite{marcelli2022purely, wesle2025near, teufel2025on} and references therein for a comparison with previous results on the Hall conductance. The common techniques from the latter works can be summarized as follows. One first identifies a \emph{non-equilibrium almost-stationary state} (NEASS), which well approximates the state of the system out of equilibrium after the application of the external electric field, if at the start the system was in a gapped equilibrium ground state (possibly modelled by a Fermi projection); the NEASS is essentially characterized by the property that it almost-commutes (up to errors $\Or(\eps^\infty)$) with the non-equilibrium Hamiltonian. This state allows to compute the current density and access the linear response coefficient, namely the Hall conductivity $\sigma\sub{Hall}$, which is by construction expressed directly in terms of the equilibrium gapped ground state by means of the so-called \emph{double-commutator formula}, in agreement with the prediction by Kubo; the fact that all further power-law corrections to Kubo's formula vanish is ultimately related to the almost-stationarity of the NEASS mentioned above.

We adopt here a mathematical setting which is similar to that in \cite{marcelli2022purely} (cf.\ below) and proceed along the line of argument sketched above. In \cite{marcelli2022purely}, the requirement that the equilibrium system be translation invariant with respect to a Bravais lattice, modelling the crystalline background, poses a (arguably unphysical) constraint on the allowed magnetic field strengths: indeed, it is required that the magnetic flux per unit cell, which is proportional to $b$ for uniform magnetic fields, be a rational multiple of the quantum of magnetic flux $\Phi_0 = h c / e$ (equal to $2\pi$ in Hartree units, which will be adopted henceforth). The aim of this paper is to show that this commensurability assumption on the magnetic field strength is indeed unnecessary: we generalize the methods (construction of the NEASS) and the results (vanishing of the power-law corrections to the response of the Hall current density beyond the linear approximation) of \cite{marcelli2022purely} to a large class of even non-uniform magnetic fields. Notice that \cite{wesle2025near} also does not assume commensurability of the magnetic field strength (cf.\ also \cite{marcelli2025gapped}), but generalizing the methods of the latter from lattice to continuum systems poses significant mathematical challenges, which are beyond the scope of the present work. On the other hand, our methods and results should be easy to adapt to the setting of one-body lattice Hamiltonians, like the ones arising from tight-binding models in condensed matter physics.

In order to construct the magnetic NEASS, we combine the recent incarnations of space-adiatic perturbation theory mentioned above, which were in turn inspired by earlier works of Nenciu \cite{nenciu1991dynamics}, with considerations inspired by \emph{gauge invariant magnetic perturbation theory}, a method developed by Nenciu, Cornean and collaborators \cite{cornean1998eigenfunction, nenciu2002asymptotic} -- see also the recent review \cite{cornean2025on}, were the connection of this method with topological transport is also discussed. Magnetic perturbation theory provides in particular exponential estimates on the off-diagonal decay of the integral kernels for certain spectral functions of magnetic Schr\"{o}dinger operators, which in turn allow to prove boundedness of nested commutators of such integral operators with the position operators (compare also \cite{denittis2017linear}); as a technical note, the latter allows to replace the smoothness of the Bloch--Floquet--Zak fibers required in \cite{marcelli2022purely}, and thus to make away with the commensurability assumption on the magnetic field strength which is required to define the Bloch--Floquet--Zak transform in the first place.

\medskip

\paragraph{\textbf{Structure of the paper.} \quad}

In Section~\ref{sec:MSO}, we introduce Bloch--Landau Hamiltonians, which function as our model for magnetic crystals, and recall their main features. Section~\ref{sec:NEASS} introduces the perturbing electric potential, and is then devoted to the construction of the magnetic NEASS as well as to the illustration of its properties. Finally, Section~\ref{sec:main} states and proves our main result on the Hall current and the validity of Kubo's formula beyond the linear response regime. The Appendices detail a number of mathematical tools which are used throughout the main text.

\medskip

\paragraph{\textbf{Acknowledgments.} \quad} 

G.\ M.\ gratefully acknowledges financial support of the European Research Council through the ERC StG MaTCh, grant agreement n.\ 101117299.

D.\ M.\ gratefully acknowledges financial support from Sapienza Università di Roma within Progetto di Ricerca di Ateneo 2023 and 2024, and from Ministero dell’Università e della Ricerca (MUR, Italian Ministry of University and Research) and Next Generation EU within PRIN 2022AKRC5P ``Interacting Quantum Systems: Topological Phenomena and Effective Theories'' and within the activities for PNRR–MUR Project no.~PE0000023-NQSTI.

\section{The unperturbed setting: Bloch--Landau operators} \label{sec:MSO}

We consider a charged quantum particle confined to a two-dimensional periodic crystal under the influence of a perpendicular magnetic field. The system at equilibrium is thus described by a magnetic Schr\"odinger operator of the following form. Define the magnetic vector potential $A\sub{lin}(x) = \frac{b}{2}(-x_2, x_1)$ in the symmetric gauge, corresponding to a constant magnetic field $B = (0, 0, b)$ perpendicular to the plane of motion. The periodic potential $V \in C^\infty(\mathbb{R}^2)$ is assumed to be bounded and translation invariant under a lattice $\Gamma = \mathbb{Z}a_1 + \mathbb{Z}a_2 \subset \mathbb{R}^2$, where $\{a_1, a_2\}$ is a basis of periodicity vectors of the underlying crystal. Corresponding to the periodic scalar potential, a further smooth periodic magnetic field could also be present: we denote by $A\sub{per} \in C^\infty(\R^2; \R^2)$ a smooth and periodic vector potential generating this magnetic field component, and finally set $A := A\sub{lin} + A\sub{per}$. A magnetic Schr\"{o}dinger operator with such periodic potentials is called a \emph{Bloch--Landau Hamiltonian}: in Hartree units, it reads
\begin{equation} \label{eqn:BLH}
H_0 := \frac{1}{2}(-\iu \nabla - A)^2 + V \quad \text{acting in } L^2(\R^2)\,.
\end{equation}

The operator $H_0$ is essentially self-adjoint on $C_0^\infty(\mathbb{R}^2)$, bounded from below, and commutes with \emph{magnetic translations}, which form a projective unitary representation of $\Gamma$:
\begin{equation} \label{eqn:Tgamma}
(T^A_\gamma \psi)(x) := \eu^{-\iu A(x) \cdot \gamma} \psi(x - \gamma), \quad \gamma \in \Gamma.
\end{equation}

We assume that $H_0$ has a spectral gap, i.e.\ there exists a compact isolated part $\sigma_0 \subset \sigma(H_0)$ which is at a positive distance $g > 0$ from the rest of the spectrum. The Fermi energy $\mu$ is chosen within this gap. The corresponding spectral projection (\emph{Fermi projection}) is defined via the Riesz formula:
\begin{equation} \label{eqn:Fermi}
\Pi_0 := \chi_{(-\infty, \mu)}(H_0) = \frac{\iu}{2\pi} \oint_\gamma (H_0 - z)^{-1} \, \di z,
\end{equation}
where $\gamma$ is a positively oriented contour in the complex energy plane enclosing $\sigma_0$. Commutativity of the Fermi projection with magnetic translations also follows by the above Riesz formula. 

\begin{remark}
We note at this stage that one could also add a bounded random potential $V_\omega$, where $\omega \in \Omega$ is a label for the disorder configuration. Provided the (family of) Hamiltonian(s) $H_\omega := H_0 + V_\omega$ is still spectrally gapped, all the considerations of this paper would go through with little modification. The typical effect of disorder is, however, to fill spectral gaps with dense pure-point spectrum \cite{aizenman1998localization, aizenman2015random}; in this situation, one refers to the presence of a \emph{mobility gap}. Generalizing the arguments of this paper to the mobility-gapped framework provides significant mathematical challenges, which we postpone to future work.
\end{remark}

The key feature we'll want to exploit is that the resolvent of a Bloch--Landau Hamiltonian, and more generally certain spectral functions of such operators, are \emph{Carleman integral operators}, whose integral kernels furthermore decay exponentially away from the diagonal. This is a standard result in the theory of (magnetic) Schrödinger operators, which can be argued using Combes--Thomas estimates \cite{combes_asymptotic_1973} in combination with the Dunford--Pettis theorem \cite{dunford1940linear}; see \cite{simon_schrodinger_1982, broderix2000continuity} and also \cite{cornean_faraday_2009, Moscolari_2023, cornean2025on} for details. We collect the results required for the following discussion in the following statement.

\begin{proposition} \label{prop:generalities}
    Let $H_0$ be the Bloch--Landau Hamiltonian defined in~\eqref{eqn:BLH}.
    \begin{enumerate}[label={(\roman*)}, ref={(\roman*)}]
        \item Fix $0<\eta<1$ and let $z \in \mathbb{C} $ satisfy $\dist(z, \sigma(H_0))=\eta>0$. Then the resolvent operator $R_{H_0}(z) = (H_0-z)^{-1}$ has an integral kernel $R_{H_0}(z; x, x')$ which is jointly continuous in $x, x' \in \R^2$, and there exist constants $C, \beta > 0$ (depending on $z$) such that
        \[ \sup_{\substack{x, x' \in \R^2 \\ \norm{x-x'}>1}} \left| \eu^{\beta \, \norm{x-x'}} \, R_{H_0}(z; x, x') \right| \le C \]
        and 
        \begin{equation} \label{eqn:L2CT}
        \sup_{x \in \R^2} \norm{R_{H_0}(z; x, \cdot) \, \eu^{\beta \, \langle \cdot - x \rangle}}_{L^2(\R^2)} =  \sup_{x \in \R^2} \norm{\eu^{\beta \, \langle \cdot - x \rangle} \, R_{H_0}(z; \cdot, x)}_{L^2(\R^2)} \le C\,.
        \end{equation}
        The above estimates and constants are uniform for $z$ ranging in compact subsets of $\set{ z \in \C : \dist(z, \sigma(H_0))=\eta>0}$. 

        The same conclusions hold for the operators
        \[ (-\iu \nabla - A)_i \, R_{H_0}(z), \quad i \in \set{1,2}\,. \]
        \item The Fermi projection $\Pi_0$, defined in~\eqref{eqn:Fermi}, admits a jointly continuous exponentially localized integral kernel $\Pi_0(x,x')$, $x, x' \in\R^2$, namely there exist constants $C, \beta > 0$ such that
        \[ \sup_{x, x' \in \R^2} \left| \eu^{\beta \, \norm{x-x'}} \, \Pi_0(x, x') \right| \le C\,. \]

        The same conclusions hold for the operators
        \[ (-\iu \nabla - A)_i \, \Pi_0, \;\; i \in \set{1,2}\,, \quad \text{and} \quad H_0 \, \Pi_0 = \Pi_0 \, H_0 \, \Pi_0. \]
    \end{enumerate}
\end{proposition}
\begin{proof}
The only statement that is not contained in the above-mentioned references is the one regarding the properties of the integral kernel for the operator $H_0 \, \Pi_0$. From the first point regarding the resolvent operator, one can deduce \cite{Moscolari_2023} that the squared resolvent operator $S_{H_0}(z) := R_{H_0}(z)^2 = (H_0-z)^{-2}$ has a jointly continuous exponentially localized integral kernel; integrating by parts the Riesz formula~\eqref{eqn:Fermi},
\[ \Pi_0 = - \frac{\iu}{2\pi} \oint_\gamma z (H_0-z)^{-2} \, \di z\,,\]
yields the same conclusion for the Fermi projection. Similarly, from
\begin{align*} 
H_0 \, \Pi_0 &= - \frac{\iu}{2\pi} \oint_\gamma z \, H_0 (H_0-z)^{-1} \, (H_0-z)^{-1} \, \di z \\
& = - \frac{\iu}{2\pi} \oint_\gamma z \, (H_0-z)^{-1} \, \di z - \frac{\iu}{2\pi} \oint_\gamma z^2 \, (H_0-z)^{-2} \, \di z\\
& = - \frac{\iu}{2\pi} \oint_\gamma \frac{z^2}{2} \, (H_0-z)^{-2} \, \di z
\end{align*}
(the last equality follows again by integration by parts on the first summand of the left-hand side), one deduces the same desired properties on the integral kernel of $H_0 \, \Pi_0$.
\end{proof}

The statements involving the magnetic momentum operators $(-\iu \nabla - A)_i$ are relevant for the applications we have in mind, as the operator
\begin{equation} \label{eqn:currentop}
J_i := \iu\,[H_0, X_i] = -\iu \partial_{x_i} - A_i, \quad i \in \set{1,2}\,,
\end{equation}
also represents the $i$-th component of the \emph{charge current operator} (for carriers of unit charge), whose response is measured in the quantum Hall effect. 

The kernel estimates provided in the previous Propositions are the ones which we will be interested in propagating and using throughout the rest of the paper. To this end, we set a general definition.

\begin{definition}[MP operators with JCEL integral kernels] \label{exp loc int ker def}
Let $T$ be a linear operator acting in $L^2(\mathbb{R}^2)$. 
\begin{itemize}
    \item We say that $T$ is \emph{magnetic periodic} (MP) if it commutes with all magnetic translation operators $T_\gamma^A$, $\gamma \in \Gamma$, as defined in~\eqref{eqn:Tgamma}. We denote by $\mathcal{L}\sub{mp}$ the set of \periodic\ operators in $L^2(\R^2)$.
    \item We say that $T$ admits a \emph{jointly continuous exponentially localized} (JCEL) \emph{integral kernel} if $T$ is an integral operator with jointly continuous kernel $T(\cdot, \cdot): \mathbb{R}^2 \times \mathbb{R}^2 \to \mathbb{C}$ and there exist positive constants $C, \beta > 0$ such that
\begin{equation}
    \sup_{x,x' \in \mathbb{R}^2} \left| \eu^{\beta\norm{x-x'}} \, T(x,x')\right| \le C\,.
\end{equation}
We will occasionally refer to $(C,\beta)$ as \emph{localization constants} for the operator $T$ and its integral kernel.
\end{itemize}

We denote by $\cS^\infty\sub{mp}$ the set of \periodic\ operators with JCEL integral kernel.
\end{definition}

Appendix~\ref{sec:ExpKern} is devoted to showing some algebraic properties of operators in $\cS^\infty\sub{mp}$ which will be used later. The condition of admitting an exponentially localized integral kernel is the ``regularity condition'' adopted in this setting, which replaces the smoothness of the Bloch--Floquet--Zak fibers required in \cite{marcelli2022purely}.

\section{Perturbed Hamiltonian and Magnetic NEASS} \label{sec:NEASS}

In the Hall effect, the crystal is pertubed by a small external electric field, and one measures the response to this perturbation of the current in the transverse direction. This Section is devoted to setting up this perturbed framework, and describing a particular class of non-equilibrium states introduced in \cite{monaco2019adiabatic, teufel_non-equilibrium_2020} to describe how the Fermi projection is ``dressed'' by such perturbation. Such a state is called \emph{non-equilibrium almost-stationary state}, or NEASS for short.

We model the perturbing electric field by means of a linear potential of small slope, and set
\begin{equation} \label{eqn:Heps}
H^\varepsilon := H_0 - \varepsilon \, X_2\,, \quad 0 < \varepsilon \ll 1\,.
\end{equation}
The construction of the NEASS for such magnetic Hamiltonians is detailed in the following Theorem: compare e.g.~\cite[Theorem~3.1]{marcelli2022purely}.

\begin{theorem} \label{thm:NEASS}
Consider the Hamiltonian $H^\varepsilon$ in \eqref{eqn:Heps}, where $H_0$ is as in \eqref{eqn:BLH}. Let $\Pi_0$ be the Fermi projection \eqref{eqn:Fermi} associated to $H_0$. Then, there exist $\eps_0 > 0$ and a sequence of operators $ \{A_j \}_{j\in \mathbb{N}} \subset \cS^\infty\sub{mp}$ such that, setting for $n \in \mathbb{N}$ and $\eps \in [0,\eps_0]$
\[S_0^\varepsilon := 0 \quad \text{and} \quad S^\varepsilon_n := \sum_{j=1}^n \varepsilon^{j-1}A_j\,,\] 
we have the following properties.
\begin{enumerate}[label={(\roman*)},ref={(\roman*)}]
    \item \label{item:NEASSi} The unitary operators
    \[ U_n^{\eps}(\lambda) := \eu^{\iu \lambda S^\eps_n}\,, \quad \eps \in [0,\eps_0]\,, \;\; \lambda \in [0,\eps]\,, \]
    are \periodic\ and such that $U_n^{\eps}(\lambda) - \Id \in \cS^\infty\sub{mp}$, with localization constants $(C_n^U, \beta_n^U)$ which are independent of $\eps$ and $\lambda$ (but possibly depend on $n$).
    \item \label{item:NEASSii} The projection    
    \begin{equation} \label{NEASS}
        \Pi_{n}^{\varepsilon} := U_n^\eps(\eps) \, \Pi_0\, U_n^\eps(\eps)^* = \eu^{\iu\varepsilon S^{\varepsilon}_n} \, \Pi_0\, \eu^{-\iu\varepsilon S^{\varepsilon}_n} \,, \quad \eps \in [0,\eps_0]\,,
    \end{equation}
    is in $\cS^\infty\sub{mp}$, with localization constants $(C_n^\Pi,\beta_n^\Pi)$ which are independent of $\eps$ (but possibly depend on $n$).
    \item \label{item:NEASSiii} There exists an operator $R_n^\eps \in \cS^\infty\sub{mp}$ such that
    \begin{equation} \label{eqn:Rneps}
    [H^\varepsilon, \Pi_n^\varepsilon] = \varepsilon^{n+1}[R^\varepsilon_n, \Pi_n^\varepsilon]\,, \quad \eps \in [0,\eps_0]\,.
    \end{equation}
    Moreover, the localization constants $(C_n^R, \beta_n^R)$ for $R_n^\eps$ can be chosen to be independent of~$\eps$ (but possibly depend on $n$). In particular, the map  $[0,\eps_0] \ni \varepsilon \mapsto R_n^\varepsilon \in \mathcal{L}\sub{mp} \cap \mathcal{B}(L^2(\mathbb{R}^2))$ is uniformly bounded. 
\end{enumerate}
\end{theorem}
\begin{proof}
If $\Pi_n^\eps$ is in the form~\eqref{NEASS}, then formally
\[ [ H^{\varepsilon},\Pi_n^{\varepsilon}] =  
 \eu^{\iu\varepsilon S^{\varepsilon}_n}\biggl[  \eu^{-\iu\varepsilon S^{\varepsilon}_n} \, H_0 \, \eu^{\iu\varepsilon S^{\varepsilon}_n} - \varepsilon \,   \eu^{-\iu\varepsilon S^{\varepsilon}_n} \, X_2 \,\eu^{\iu\varepsilon S^{\varepsilon}_n}, \Pi_0\biggr] \eu^{-\iu\varepsilon S^{\varepsilon}_n}
\]
Hence, we have to choose the operators $A_j$ in such a way that there exists $R^{\varepsilon}_n$ with the properties stated in the Theorem such that 
\begin{equation}
   \label{eq: step1} \biggl[  \eu^{-\iu\varepsilon S^{\varepsilon}_n} \, H_0 \, \eu^{\iu\varepsilon S^{\varepsilon}_n} - \varepsilon \, \eu^{-\iu\varepsilon S^{\varepsilon}_n} \, X_2 \, \eu^{\iu\varepsilon S^{\varepsilon}_n}, \Pi_0\biggr] = \varepsilon^{n+1} \biggl[\eu^{-\iu\varepsilon S^{\varepsilon}_n} \, R^{\varepsilon}_n \, \eu^{\iu\varepsilon S^{\varepsilon}_n}, \Pi_0 \biggr].
\end{equation}

Let $B$ be a short-hand notation for $H_0$ or $X_2$. Consider the Taylor expansion in $\lambda$ near $\lambda_0 = 0$ of the function 
\[
\lambda \mapsto \eu^{-\iu\lambda S^{\varepsilon}_n} \, B \, \eu^{\iu\lambda S^{\varepsilon}_n}\,.
\]
Evaluating in $\lambda = \varepsilon$, we get, for some $\tilde{\varepsilon} \in [0, \varepsilon]$ 
\[
    \eu^{-\iu \varepsilon S_n^\varepsilon} \, B \, \eu^{\iu \varepsilon S_n^\varepsilon} = \sum_{k=0}^{n} \frac{\varepsilon^k}{k!} \,\mathcal{L}_{S_n^\varepsilon}^k (B) + \frac{\varepsilon^{n+1}}{(n+1)!} \, \eu^{-\iu \tilde{\varepsilon} S_n^\varepsilon} \,\mathcal{L}_{S_n^\varepsilon}^{n+1} (B) \, \eu^{\iu \tilde{\varepsilon} S_n^\varepsilon} \,,
\]
where we employed the notation $\mathcal{L}_A(B) := - \iu [A,B]$ for the Liouvillian superoperator of commutator with $-\iu\,A$, and correspondingly $\mathcal{L}_A^k$ for the $k$-nested commutator with $-\iu \, A$.

We now spell out the commutators in the sum, using the \textit{ansatz} $ S_n^\varepsilon = \sum_{j=1}^n \varepsilon^{j-1}A_j$: 
\begin{align*}
   \eu^{-\iu \varepsilon S_n^\varepsilon} \, B \, \eu^{\iu \varepsilon S_n^\varepsilon} & = B + \sum_{k=1}^n \frac{(-\iu)^k}{k!} \, \sum_{\substack{j = (j_1,\ldots,j_k) \in \mathbb{N}^k\\ 1 \le j_i \le n}} \varepsilon^{|j|} \, [A_{j_1},[A_{j_2},\dots[A_{j_k},B]\dots]] \\
   & \quad + \frac{\varepsilon^{n+1}}{(n+1)!} \, \eu^{-\iu \tilde{\varepsilon} S_n^\varepsilon} \,\mathcal{L}_{S_n^\varepsilon}^{n+1} (B) \, \eu^{\iu \tilde{\varepsilon} S_n^\varepsilon}
\end{align*}
where $|j|= \sum_i j_i \in \mathbb{N}$ for $j = (j_1, j_2, \dots, j_k) \in \mathbb{N}^k$.

Next, we collect the terms corresponding to the various powers of $\varepsilon$ and retain explicitly the ones with order $m \le n$: the above expression results in
\[
\eu^{-\iu \varepsilon S_n^\varepsilon} \, B \, \eu^{\iu \varepsilon S_n^\varepsilon} = \sum_{m=0}^n \varepsilon^m B_m + \varepsilon^{n+1}B_{n+1}(\varepsilon)\,,
\]
where 
\begin{equation}\label{eq:Bm} 
B_0 := B\,, \quad B_m := \sum_{k=1}^{m} \frac{(-\iu)^k}{k!} \sum_{\substack{j \in \mathbb{N}^{k}\\ 1 \le j_i \le m-k+1\\ |j|=m }} [A_{j_1},[A_{j_2},\dots[A_{j_k},B]\dots]] \text{ for } 0 < m \le n\,,
\end{equation}
and
\begin{equation} \label{eqn:Bn+1}
\begin{aligned}
B_{n+1}(\eps) & := \frac{1}{(n+1)!} \, \eu^{-\iu \tilde{\varepsilon} S_n^\varepsilon} \,\mathcal{L}_{S_n^\varepsilon}^{n+1} (B) \, \eu^{\iu \tilde{\varepsilon} S_n^\varepsilon} \\
& \quad + \sum_{k=2}^{n} \frac{(-\iu)^k}{k!} \sum_{\substack{j \in \mathbb{N}^{k} \\ 1 \le j_i \le n\\ |j|\ge n+1}} \eps^{|j|-(n+1)} \, [A_{j_1},[A_{j_2},\dots[A_{j_k},B]\dots]]
\end{aligned}
\end{equation}
We observe that the term $B_m$, $0 < m \le n$, depends only on nested commutators (of which there are at most $m$) with the operators $A_j$ for $j\le m$.

We apply the previous expansion to $B=H_0$ and, up to order $n-1$, to $B=X_2$. We denote by ${(H_0)}_{m}$ and ${(X_2)}_m$ the respective coefficients of the expansions. We substitute into \eqref{eq: step1}, noting that the presence of an extra factor $\varepsilon$ in the perturbation will shift the indices all the coefficients in the expansion for $X_2$:
\begin{equation} \label{eqn:reminder}
    \sum_{m=1}^n \varepsilon^m \bigl[ {(H_0)}_{m} - {(X_2)}_{m-1}, \Pi_0 \bigr] + \varepsilon^{n+1} \bigl[ {(H_0)}_{n+1}(\varepsilon) - {(X_2)}_{n}(\varepsilon), \Pi_0 \bigr] = \varepsilon^{n+1} \biggl[\eu^{-\iu\varepsilon S^{\varepsilon}_n} R^{\varepsilon}_n \eu^{\iu\varepsilon S^{\varepsilon}_n}, \Pi_0 \biggr]
\end{equation}
Thus, it is sufficient to determine $A_1, A_2, \dots , A_n$ in such a way that, for all $1\le m\le n$, \begin{equation} \label{eq:ric}
    \bigl[ {(H_0)}_{m} - {(X_2)}_{m-1}, \Pi_0 \bigr] = 0\,.
\end{equation}
The previous considerations suggest the chance of recursively computing the~$A_j$'s. To this end, we rewrite 
\[{(H_0)}_{m}= \mathcal{L}_{A_m}(H_0) + L_{m-1} = -\mathcal{L}_{H_0}(A_m) + L_{m-1}\] 
where $L_{m-1}$ collects all the commutators between $H_0$ and $A_\mu$ for $\mu < m$, and set up a recursive process so that if the latter $A_\mu$'s are known then the above equation can be solved for $A_m$. 

We begin by computing $A_1$ from \eqref{eq:ric}. For $m=1$ we get that $L_0 = 0$ and ${(X_2)}_0 = 0$, meaning that the equation for $A_1$ we want to solve is
\[ 0= [-\mathcal{L}_{H_0}(A_1) - X_2, \Pi_0].\] 
Since $[H_0, \Pi_0]=0$, we find that 
\[
[\mathcal{L}_{H_0}(A_1),\Pi_0]= \mathcal{L}_{H_0}([A_1,\Pi_0]) - [A_1,\mathcal{L}_{H_0}(\Pi_0)] = \mathcal{L}_{H_0}([A_1,\Pi_0])
\]
hence the equation to solve for $A_1$ is equivalently written as 
\[ \mathcal{L}_{H_0}([A_1,\Pi_0]) = - [X_2,\Pi_0].\] 
We note that the operator $[X_2,\Pi_0]$ is off-diagonal with respect to the decomposition induced by $\Pi_0$ (compare Appendix~\ref{sec:OD}) and in addition it is in $\cS^\infty\sub{mp}$ in view of Proposition~\ref{estimates for commutators}. By Proposition~\ref{invlio}, the inverse Liouvillian $\mathcal{L}^{-1}_{H_0}$ is well-defined on $[X_2,\Pi_0]$ due to these properties, and we can set
\[[A_1, \Pi_0]= \mathcal{L}_{H_0}^{-1}(-[X_2, \Pi_0])\,.\]
Taking an additional commutator with $\Pi_0$ we conclude with Proposition~\ref{4.1} that
\[A_1\su{OD} = [[A_1, \Pi_0],\Pi_0] = [\mathcal{L}_{H_0}^{-1}(-[X_2, \Pi_0]), \Pi_0]= - \mathcal{L}_{H_0}^{-1}(X_2\su{OD})\,. \]
We have therefore uniquely determined the off-diagonal part of  $A_1$. We choose $A_1\su{D}:=0$ so that we can define \begin{equation}
    \label{eq:A1} A_1 := A_1\su{OD} = - \mathcal{L}_{H_0}^{-1}(X_2\su{OD})
\end{equation}
and $A_1$ is in $\cS\sub{mp}^\infty$ in view of Proposition~\ref{invlio}\ref{item:ILiii}. 

For $1<m\le n$, we require that \[0= -\mathcal{L}_{H_0}([A_m, \Pi_0])+[L_{m-1},\Pi_0] - [{(X_2)}_{m-1},\Pi_0]\]
which, proceeding as for $A_1$, we rewrite as
\[ [A_m, \Pi_0] = \mathcal{L}_{H_0}^{-1}([L_{m-1}-{(X_2)}_{m-1},\Pi_0])\,.\]
Commuting with $\Pi_0$ both sides, we obtain 
\[
A_m\su{OD} = \mathcal{L}_{H_0}^{-1}((L_{m-1}-{(X_2)}_{m-1})\su{OD})\,.
\] 
We denote $D_{m-1} := (L_{m-1}-{(X_2)}_{m-1})\su{OD} $ and observe that it is determined by $A_1, \dots , A_{m-1}$, which are assumed to have been already computed and in $\cS\sub{mp}^{\infty}$. Therefore, we again choose $A_m\su{D} := 0$, so that  $A_m$ is purely off-diagonal and in $\cS\sub{mp}^{\infty}$ as well. Thus we can finally define 
\begin{equation} \label{eqn:Deps} 
S_n^\varepsilon := \sum_{j=1}^n \varepsilon^{j-1}A_j = \mathcal{L}^{-1}_{H_0}(D^\eps_n) \in \cS^\infty\sub{mp}\,, \quad D^\eps_n :=\sum_{l=0}^{n-1}\varepsilon^l D_l \in \cS^\infty\sub{mp}\,.
\end{equation}
Notice that, since $S_n^\eps$ is a finite sum of the $A_j$'s for fixed $n \in \N$, the localization constants of $S_n^\eps$ can be bounded uniformly in $\eps \in [0,\eps_0]$ for any $\eps_0>0$.

Having defined the $A_j$'s and $S_n^\eps$, we can now prove the remaining statements in~\ref{item:NEASSi} and~\ref{item:NEASSii}. For $\lambda \in [0,\eps]$, define $F(\lambda,x) := \sum_{k\ge1} (\iu\lambda x)^k/k!$. Then it follows from Proposition~\ref{prop exp loc int ker}\ref{item:Tiii} that
\[
   U_n^\eps(\lambda) = \eu^{\iu\lambda S^\eps_n} = \Id + F(\lambda, S^\eps_n), \quad F(\lambda,S^\eps_n)\in \cS^\infty\sub{mp}\,.
\]
Again from Proposition~\ref{prop exp loc int ker}\ref{item:Tiii}, for fixed $n \in \N$ the localization constants of $U_n^\eps(\lambda)-\Id = F(\lambda,S_n^\eps)$ can be chosen in a uniform way in $\lambda \in [0,\eps]$ and $\eps \in [0, \eps_0]$, possibly at the expense of making $\eps_0$ smaller.

To see that the NEASS $\Pi^\varepsilon_n$ is in $\cS^\infty\sub{mp}$, we write it as a sum of products of operators in $\cS^\infty\sub{mp}$ as follows:
\begin{align*}
    \Pi^\eps_n &= U_n^\eps(\eps)\, \Pi_0 \, U_n^\eps(\eps)^* = \big[\Id + F(\eps,S^\eps_n)\big] \, \Pi_0 \, \big[ \Id + F(\eps,S^\eps_n)\big]^* \\
    &= \Pi_0 + F(\eps,S^\eps_n)\, \Pi_0 + \Pi_0 \,F(\eps,S^\eps_n)^* + F(\eps,S^\eps_n) \, \Pi_0 \, F(\eps,S^\eps_n)^*\,. 
\end{align*} 
The statement on the localization constants of $\Pi_n^\eps$ also follows from the above rewriting and the previous observation on the localization constants of $F(\eps,S_n^\eps)$.

Finally, coming back to~\eqref{eqn:reminder}, we see that we can set
\[ R_n^\eps := U_n^\eps(\eps) \, (H_0)_{n+1}(\eps) \, U_n^\eps(\eps)^* - U_n^\eps(\eps) \, (X_2)_n(\eps) \, U_n^\eps(\eps)^* \,. \]
As is apparent from~\eqref{eqn:Bn+1}, in view of the previous considerations, for fixed $n \in \N$ the operators $(H_0)_{n+1}(\eps)$ and  $(X_2)_n(\eps)$ are in $\cS^\infty\sub{mp}$ with localization constants which can be chosen to be independent of $\eps \in [0,\eps_0]$. By the previous argument for $\Pi_n^\eps$, conjugation by the unitary $U_n^\eps(\eps)$ does not spoil this property.
This concludes the proof.
\end{proof}

\begin{corollary} \label{cor:funzioni di Pieps}
Consider the NEASS $\Pi^\varepsilon_n \in \cS\sub{mp}^\infty$ defined in \eqref{NEASS}. The following operators are in $\cS\sub{mp}^\infty$ as well:
\begin{enumerate}[label={(\roman*)},ref={(\roman*)}]
    \item \label{item:PiepsX} $[\Pi^\varepsilon_n, X_j]$ for $j\in \set{1,2}$;
    \item \label{item:H0Pieps} $H_0 \, \Pi^\eps_n$;
    \item \label{item:JPieps} $(-\iu\nabla-A)_i \, \Pi^\varepsilon_n$ for $i \in \set{1,2}$;
\end{enumerate}
For all such operators, the localization constants can be chosen to be independent of $\eps \in [0,\eps_0]$ (but possibly depend on $n \in \N$).
\end{corollary}
\begin{proof}
Point \ref{item:PiepsX} follows immediately from Proposition~\ref{estimates for commutators}. As for \ref{item:H0Pieps}, let us write, with the notation of the previous proof,
\begin{equation} \label{eqn:H0Peps}
\begin{aligned}
H_0 \, \Pi_n^\eps & = H_0 \, \Pi_0 \, \eu^{-\iu \eps S_n^\eps} + H_0 \, F(\eps,S_n^\eps) \, \Pi_0 \, \eu^{-\iu \eps S_n^\eps} \\
& = H_0 \, \Pi_0 \, \eu^{-\iu \eps S_n^\eps} + F(\eps,S_n^\eps) \, H_0 \, \Pi_0 \, \eu^{-\iu \eps S_n^\eps} + [H_0, F(\eps,S_n^\eps)] \, \Pi_0 \, \eu^{-\iu \eps S_n^\eps} \\
& = H_0 \, \Pi_0 + H_0 \, \Pi_0 \, F(\eps,S_n^\eps)^* + F(\eps,S_n^\eps) \, H_0 \, \Pi_0 + F(\eps,S_n^\eps) \, H_0 \, \Pi_0 \, F_\eps(S_n^\eps)^* \\
& \quad + [H_0, F(\eps,S_n^\eps)] \, \Pi_0 + \, [H_0, F(\eps,S_n^\eps)] \, \Pi_0 \, F(\eps,S_n^\eps)^*\,.
\end{aligned}
\end{equation}
Since
\begin{align*} 
[H_0,F(\eps,S^\eps_n)] & = \iu \, \mathcal{L}_{H_0}\left(F(\eps,S^\eps_n)\right) = \iu \sum_{k \ge 1} \frac{(-\iu \eps)^{k}}{k!} \, \mathcal{L}_{H_0}\left( \mathcal{L}_{H_0}^{-1}(D^\eps_n)^{k} \right) \\
& = \iu \sum_{k \ge 1} \frac{(-\iu \eps)^{k}}{k!} \, \sum_{j=1}^{k} \mathcal{L}_{H_0}^{-1}(D^\eps_n)^{j-1} \, D_n^\eps \,  \mathcal{L}_{H_0}^{-1}(D^\eps_n)^{k-j} \,,
\end{align*}
\eqref{eqn:H0Peps} exhibits $H_0 \, \Pi_n^\eps$ as a linear combination of elements of $\cS^\infty\sub{mp}$, and thus establishes this operator as in $\cS^\infty\sub{mp}$ itself.

Statement \ref{item:JPieps} follows from a similar argument, writing
\[
(-\iu\nabla-A)_i \, \Pi^\varepsilon_n = (-\iu\nabla-A)_i \, (H_0 - \iu)^{-1} \, (H_0 - \iu) \, \Pi_n^\eps 
\]
and using the previous point and the properties stated in Proposition~\ref{prop:generalities}.
\end{proof}

\section{Validity of Kubo's Formula beyond the linear response regime} \label{sec:main}

In this Section we return to the problem of computing the Hall current in the NEASS constructed in Theorem~\ref{thm:NEASS}. In the context of infinite, translation-invariant systems, the correct functional to take expectations of extensive observables in extended states is the \emph{trace per unit area} (or per unit volume, in higher dimensions), which we denote by $\mathcal{T}$: we recall the main properties of this functional and of operators with finite trace per unit area (whose class we denote by $\cS^1\sub{mp}$) in Appendix~\ref{sec:TPUA}. 

The main property we'll exploit in this Section, stated as Proposition\ref{prop classe traccia puv}\ref{item:Tinf}, is that operators $T \in \cS\sub{mp}^\infty$ do have finite trace per unit area, computed by the expression
\[ \mathcal{T}(T) = \frac{1}{|\Lambda_1|} \int_{\Lambda_1} T(x,x) \, \di x \]
where $\Lambda_1$ denotes the unit cell of the periodicity lattice $\Gamma$ (compare~\eqref{eqn:Lambda1}). In particular, Proposition~\ref{prop:generalities} implies that $\Pi_0 \in \cS\sub{mp}^\infty$ has a finite \emph{integrated density of states} $\mathcal{T}(\Pi_0)$; moreover, Corollary~\ref{cor:funzioni di Pieps}\ref{item:JPieps} together with the definition~\eqref{eqn:currentop} of the current operator yields that the Hall current density
\[ j\sub{Hall}(\eps) := \mathcal{T}\left(J_1 \, \Pi^\eps_n\right) \]
is well posed. We are now able to state the main result (compare \cite[Theorem~4.1]{marcelli2022purely}).

\begin{theorem}  \label{teo:4.2.2}
Let $H_0$ be as in \eqref{eqn:BLH}, $H^\varepsilon = H_0 - \varepsilon X_2$ be as in \eqref{eqn:Heps}, and, for $n \in \mathbb{N}$, let $\Pi^\eps_n$ be as in Theorem \ref{thm:NEASS}. Let also $J_1$ denote the current operator defined in \eqref{eqn:currentop}. Then
\[ j\sub{Hall}(\eps) = \varepsilon \,\sigma\sub{Hall} + \mathcal{O}(\varepsilon^{n+1}) \]
where 
\begin{equation} \label{sigmahall} 
\sigma\sub{Hall}:= \iu \, \mathcal{T}\big(\Pi_0\,\big[[\Pi_0, X_1],[\Pi_0, X_2]\big]\,\Pi_0\big)\,.
\end{equation}
\end{theorem}

The above theorem states that the linear response coefficient of the current operator $J_1$ is given by the Hall conductivity $\sigma_{\text{Hall}}$, which is defined exclusively through the equilibrium Fermi projection $\Pi_0$, and moreover that all power-law, higher-order corrections in $\varepsilon$ to this linear response term vanish. This establishes, in the present context, the validity of Kubo's formula beyond linear response.

The proof of the theorem is based on a sequence of intermediate steps, stated as the following Propositions.

\begin{proposition} \label{4.2.4}
    Let $P$ be a \periodic\ projection on $L^2(\mathbb{R}^2)$  which admits a JCEL integral kernel, in the sense of Definition \ref{exp loc int ker def}. Let $A \in \mathcal{L}_{\text{mp}}$ be an operator such that $PAP$ has a JCEL integral kernel. Then, for any $j \in \{1,2\}$, it holds that $[PAP,PX_j P] \in \cS^1\sub{mp}$ and
    \begin{equation} \label{eqn:PAPXP}
        \mathcal{T}([PAP, PX_j P]) = 0\,.
    \end{equation}
\end{proposition}
\begin{proof}
    We begin by noting that $P \in \cS^1\sub{mp}$ in view of Proposition~\ref{prop classe traccia puv}~\ref{item:Tinf}, and that
    \[ [PAP, PX_j P] = [PAP, X_j] - [PAP, X_j\su{OD}] \]
    where $X_j\su{OD}$ refers to the off-diagonal part of the operator $X_j$ with respect to the decomposition of $L^2(\R^2)$ induced by $P$. By Propositions~\ref{prop exp loc int ker}~\ref{item:Tii} and~\ref{estimates for commutators}, both summands are in $\cS^\infty\sub{mp}$, and therefore in $\cS^1\sub{mp}$ again by Proposition~\ref{prop classe traccia puv}~\ref{item:Tinf}. With $T := P A P$, the operator $[T, X_j]$ has integral kernel
    \[ [T, X_j](x,x') = (x_j'-x_j) \, T(x,x') \]
    which vanishes on the diagonal $x=x'$; therefore, its trace per unit area~\eqref{traccia puv su lambda 1} vanishes as well. Besides, the trace of $[T,X_j\su{OD}]$ is also zero by cyclicity of the trace per unit area, Proposition~\ref{prop classe traccia puv}~\ref{item:Tcyclic}. The conclusion~\eqref{eqn:PAPXP} follows.
\end{proof}

\begin{remark} \label{remark: 4.2.5}
The above Proposition does \emph{not} apply in the case where $A = X_i$ for $i\neq j$, since it is not \periodic. Nonetheless, the operator $[PX_iP, PX_jP]$ and its trace per unit area are still of interest. Let us compute
\begin{align*} 
[PX_iP, PX_jP] & = [P(X_i - X_i\su{OD})P, P(X_j - X_j\su{OD}) P] \\
& = P \left[ X_i - X_i\su{OD}, X_j - X_j\su{OD}\right] P \\
& = P \left( - \left[ X_i\su{OD}, X_j\right] - \left[ X_i , X_j\su{OD}\right] + \left[ X_i\su{OD}, X_j\su{OD}\right] \right) P\,.
\end{align*}
This equality also exhibits the operator $[PX_iP, PX_jP]$ as in $\cS^\infty\sub{mp} \subset \cS^1\sub{mp}$. Only the diagonal part of an operator contributes to its trace per unit area: from this consideration and Proposition~\ref{4.1}~\ref{item:ABOD}, we conclude that
\[ \mathcal{T}\left(P\left[ X_i\su{OD}, X_j\right]P\right) = \mathcal{T}\left(P\left[ X_i\su{OD}, X_j\su{OD}\right]P\right) = \mathcal{T} \left(P\left[ X_i , X_j\su{OD}\right]P\right) \]
and therefore that
\[ \mathcal{T}([PX_iP, PX_jP]) = - \mathcal{T} \left( P \left[ X_i\su{OD}, X_j\su{OD}\right] P \right) = - \mathcal{T} \left(P[[P,X_i],[P,X_j]]P\right) \]
where the last equality follows from Proposition~\ref{4.1}~\ref{item:AOD} by an immediate check. The operator $[PX_iP, PX_jP]$, however, is \emph{not} the commutator of operators with finite trace per unit area, and as such one cannot invoke cyclicity of the trace per unit area to conclude that $\mathcal{T}([PX_iP, PX_jP])=0$. Indeed, for $P=\Pi_0$ the Fermi projection, $i=1$ and $j=2$, Equation~\eqref{sigmahall} establishes this trace per unit area as proportional to the Hall conductivity.
\end{remark}

\begin{proposition}[Chern-Simons formula] \label{prop:CS}
    Let $P$ be a \periodic\ projection on $L^2(\mathbb{R}^2)$ that admits a JCEL integral kernel, in the sense of Definition~\ref{exp loc int ker def}. Let $U \in \mathcal{U}(L^2(\R^2))$ be a \periodic\ unitary such that $U-\mathds{1}$ admits a JCEL integral kernel. Define $P_U := UPU^{-1} \in \cS^\infty\sub{mp}$. Then
    \begin{equation}
        \mathcal{T}([P_U X_i P_U, P_U X_j P_U]) = \mathcal{T}([PX_i P, PX_j P])\,.
    \end{equation}
\end{proposition}
\begin{proof}
We write
\begin{equation} \label{3012}
\begin{aligned}
    U^{-1}&[P_U X_i P_U, P_U X_j P_U]U = \\
        &= [P U^{-1} X_i UP, PU^{-1}X_jUP] = \\ &= 
        [PX_i P, PX_j P] + [P U^{-1} [X_i, U]P, PX_jP]+ \\ &\quad + [PX_iP, PU^{-1}[X_j, U]P] + [PU^{-1}[X_i, U]P, PU^{-1}[X_j, U]P]\,.
\end{aligned}
\end{equation}
Notice that $U^{-1} - \Id = U^* - \Id = (U-\Id)^*$ is \periodic\ and has a JCEL integral kernel by Proposition~\ref{prop exp loc int ker}~\ref{item:Ti}, and therefore also
\[ U^{-1}[X_i, U] = [X_i, U-\Id] + (U^{-1} - \Id) \, [X_i, U-\Id] \]
and $U^{-1}[X_j, U]$ are in $\cS^\infty\sub{mp}$ also in view of Proposition~\ref{estimates for commutators}. Thus, Propositions~\ref{4.2.4} and~\ref{prop classe traccia puv}~\ref{item:Tcyclic} give that the last three summand on the right-hand side of the above expression~\ref{3012} have vanishing trace per unit area. 

The proof is complete in view of the invariance of the trace per unit area under (appropriate) unitary conjugation, Proposition~\ref{prop:T(UTU-1)}.
\end{proof}

\begin{proof}[Proof of Theorem~\ref{teo:4.2.2}]
Let us consider the orthogonal splitting of $L^2(\R^2)$ induced by $\Pi^\eps_n$, and denote correspondingly by $T^{\text{D}_\eps}$ (respectively $T^{\text{OD}_\eps}$) the diagonal (respectively off-diagonal) parts of an operator $T$ in $L^2(\R^2)$ (compare Definition~\ref{def:DOD}). By~\eqref{NEASS} and Proposition~\ref{4.1}~\ref{item:AOD}, we observe that
\[ H^\eps = \left(H^\eps\right)^{\text{D}_\eps} + \left(H^\eps\right)^{\text{OD}_\eps} = \left(H^\eps\right)^{\text{D}_\eps} + \big[ [H^\eps,\Pi^\eps_n], \Pi^\eps_n \big] = \left(H^\eps\right)^{\text{D}_\eps} + \eps^{n+1} \, \left(R^\eps_n\right)^{\text{OD}_\eps}\,.\]
Using the above relation and the cyclicity of the trace per unit area, Proposition~\ref{prop classe traccia puv}~\ref{item:Tcyclic}, the following then holds:
\begin{equation} \label{eqn:computation}
\begin{aligned}
-\iu \, \mathcal{T}\left(J_1 \, \Pi_n^{\varepsilon}\right) &= \mathcal{T}\left(\Pi_n^{\varepsilon}[H^{\varepsilon}, X_1]\Pi_n^{\varepsilon}\right) \\
&= \mathcal{T}\left(\left[\Pi_n^{\varepsilon}H^{\varepsilon}\Pi_n^{\varepsilon}, \Pi_n^{\varepsilon}X_1\Pi_n^{\varepsilon}\right]\right)  + \varepsilon^{n+1}\, \mathcal{T}\left(\Pi_n^{\varepsilon} \, [(R_n^{\varepsilon})^{\text{OD}_\eps},X_1] \, \Pi_n^{\varepsilon} \right) \,.
\end{aligned}
\end{equation}
The coefficient of $\eps^{n+1}$ on the right-hand side of the above equality is uniformly bounded in $\eps$. Indeed, setting $T_n^\eps := \big[ [R_n^\eps,\Pi^\eps_n], \Pi^\eps_n \big]$, Theorem~\ref{thm:NEASS} implies that $T_n^\eps \in \cS^\infty\sub{mp}$, and that its localization constants $(C_n^T, \beta_n^T)$ are uniformly bounded in $\eps$. Then
\begin{align*} 
\left| \mathcal{T}\left(\Pi_n^{\varepsilon} \, [T_n^{\varepsilon},X_1] \, \Pi_n^{\varepsilon} \right) \right| & = \frac{1}{|\Lambda_1|}\,\left| \int_{\Lambda_1} \di x \int_{\R^2} \di y \,\Pi_n^{\varepsilon}(x,y) \, (y_1-x_1) \, T_n^{\varepsilon}(y,x) \right| \\
& \le C_n^\Pi \, C_n^T \int_{\R^2} \di y \, |y| \eu^{-(\beta_n^\Pi + \beta_n^T) |y|}
\end{align*}
is uniformly bounded in $\eps$, as wanted.

We resume the computation in~\eqref{eqn:computation} as
\[ -\iu \, \mathcal{T}\left(J_1 \, \Pi_n^{\varepsilon}\right) = \mathcal{T}\left(\left[\Pi_n^{\varepsilon}H_0\Pi_n^{\varepsilon}, \Pi_n^{\varepsilon}X_1\Pi_n^{\varepsilon}\right]\right) - \eps \, \mathcal{T}\left(\left[\Pi_n^{\varepsilon}X_2\Pi_n^{\varepsilon}, \Pi_n^{\varepsilon}X_1\Pi_n^{\varepsilon}\right]\right) + \mathcal{O}(\eps^{n+1})\,.\]
From Corollary~\ref{cor:funzioni di Pieps}~\ref{item:H0Pieps}, we know that $\Pi_n^{\varepsilon}H_0\Pi_n^{\varepsilon}$ is in $\cS^\infty\sub{mp}$; therefore, Proposition~\ref{4.2.4} implies that the first term on the right-hand side of the above is zero. On the other hand, by Remark~\ref{remark: 4.2.5} and Proposition~\ref{prop:CS} the second summand can be rewritten as
\[ \mathcal{T}([\Pi_n^{\varepsilon}X_2\Pi_n^{\varepsilon},\Pi_n^{\varepsilon}X_1\Pi_n^{\varepsilon}]) = \mathcal{T}([\Pi_0X_2\Pi_0,\Pi_0X_1\Pi_0]) = -\mathcal{T}([\Pi_0[[\Pi_0,X_2],[\Pi_0, X_1]]\Pi_0)\,. \]
Thus, we have concluded the proof.
\end{proof}

\appendix

\section{Exponentially localized integral kernels} \label{sec:ExpKern}

This Appendix formalizes the class of operators in $\cS^\infty\sub{mp}$ (compare Definition~\ref{exp loc int ker def}), providing the basic analytic and algebraic tools needed in the main text.

\begin{remark} \label{rmk:Schur}
An operator with a (not necessarily jointly continuous but measurable) exponentially localized integral kernel is necessarily a bounded operator from $L^2(\R^2)$ to itself, and moreover its norm is bounded by its first localization constant $C$. This follows at once from Schur's test.
\end{remark}

\begin{proposition}\label{prop exp loc int ker}
Let $T_1$, $T_2$, and $T$ admit JCEL integral kernels with localization constants $(C_1, \beta_1)$, $(C_2, \beta_2)$, and $(C, \beta)$, respectively. Then:
\begin{enumerate}[label={(\roman*)},ref={(\roman*)}]
\item \label{item:Ti} for any $\alpha \in \C$, $T_1 + \alpha T_2$ admits a JCEL integral kernel with localization constants $(C_1 + |\alpha| \, C_2, \beta_1 + \beta_2)$; $T^*$ admits a JCEL integral kernel with localization constants $(C,\beta)$;
\item \label{item:Tii} $T_1 T_2$ admits a JCEL integral kernel with localization constants $(C_1 C_2, \beta')$ for any $\beta'$ such that $0 < \beta' < \min\{\beta_1, \beta_2\}$;
\item  \label{item:Tiii} If $F(x) = \sum_{n \geq 1} a_n x^n$ is an analytic function with radius of convergence $r > C$, then $F(T)$ admits a JCEL integral kernel with localization constants $(\sum_{n\ge 1}|a_n| \, C^n, \beta')$ for any $\beta'$ such that $0 < \beta' < \beta$.
\end{enumerate}
\end{proposition}
\begin{proof}
\ref{item:Ti} follows immediately from the definitions.

As for \ref{item:Tii}, the kernel of $T_1 T_2$ is given by the integral
\[
(T_1 T_2)(x, x') = \int \di y\, T_1(x, y)\, T_2(y, x').
\]
Denote by $(C_1,\beta_1)$ and $(C_2,\beta_2)$ the localization constants of $T_1$ and $T_2$, respectively. Pick $0 < \beta' < \min\{\beta_1, \beta_2\}$. Then
\begin{align*}
\left| \eu^{\beta' \, \norm{x-x'}} \, (T_1 T_2)(x, x') \right| & \le  \int \di y\, \left| \eu^{\beta' \, \norm{x-y}} \, T_1(x, y) \right| \, \left| \eu^{\beta' \, \norm{y-x'}} \, T_2(y, x') \right| \\
& \le \left( \int \di y\, \eu^{-2(\beta_1-\beta') \, \norm{x-y}} \, \left| \eu^{\beta_1 \, \norm{x-y}} \, T_1(x, y) \right|^2 \right)^{1/2} \\
& \qquad \left( \int \di y\, \eu^{-2(\beta_2-\beta') \, \norm{y-x'}} \, \left| \eu^{\beta_2 \, \norm{y-x'}} \, T_2(y, x') \right|^2 \right)^{1/2} \\
& \le C_1 \, C_2 \, \left(\int \di y \, \eu^{-(\beta_1 - \beta') \, \norm{y}} \right)^{1/2} \, \left(\int \di y \, \eu^{-(\beta_2 - \beta') \, \norm{y}} \right)^{1/2} 
\end{align*}
which is bounded uniformly in $x,x'\in \R^2$. Thus, $T_1\,T_2$ admits an exponentially localized integral kernel.

To prove joint continuity, assume that $x_n \to x \in \R^2$ and $x_n' \to x' \in \R^2$ as $n \to \infty$. Then
\begin{equation} \label{eqn:T1T2}
\begin{aligned}
\big| (T_1 T_2)(x_n, x_n') & - (T_1 T_2)(x, x') \big| \le \int \di y\, \left| T_1(x_n, y)\, T_2(y, x_n') - T_1(x, y)\, T_2(y, x') \right| \\
& \le\int \di y\, \left| T_1(x_n, y) \right| \, \left| T_2(y, x_n') - T_2(y,x')\right| \\
& \qquad + \int \di y \, \left|T_1(x_n,y) - T_1(x, y)\right|\, \left| T_2(y, x') \right| \,.
\end{aligned}
\end{equation}
Let us show that one can pass the limit $n \to \infty$ under the integrals by dominated convergence. To this end, let us first choose $M>0$ such that the sequence $\norm{x_n-x}$, which converges to 0, is bounded by $M$ uniformly in $n$. Then we have
\[ \left|T_1(x_n,y) - T_1(x, y)\right| \le 2 \, C_1\,, \quad \left| T_2(y, x_n') - T_2(y,x')\right| \le 2 \, C_2\,, \quad \left| T_2(y, x') \right| \le C_2 \, \eu^{-\beta_2 \, \norm{y-x'}} \]
and also
\[ \left| T_1(x_n, y) \right| \le C_1 \, \eu^{-\beta_1 \, \norm{x_n-y}} \le C_1 \, \eu^{-\beta_1 \, \left( \norm{x-y} - \norm{x_n-x} \right)} \le C_1 \, \eu^{\beta_1 \, M} \, \eu^{-\beta_1 \, \norm{x-y}} \]
where in the last step we used the reverse triangle inequality 
\[ \norm{x_n - y} \ge \big| \norm{x_n-x} - \norm{x-y} \big| \ge \norm{x_n-x} - \norm{x-y}.\]
The conclusion follows.

\ref{item:Tiii} (Compare also \cite[Lemma~3.5 and Lemma~A.1]{cornean2025on}.) Since $T^n$ is a product of operators with JCEL integral kernels, it admits a JCEL integral kernel by iterating \ref{item:Tii}; its localizations constants are $(C^n, \beta')$ for any $0<\beta'<\beta$. The analytic functional calculus $F(T) = \sum_{n \geq 1} a_n T^n$ defines an absolutely convergent series in view of the norm bound $\|T\| \le C < r$, see Remark~\ref{rmk:Schur}. Hence, $F(T)$ has an integral kernel
\[ F(T)(x,x') = \sum_{n \geq 1} a_n \, T^n(x,x')\,. \]
As an absolutely convergent series of jointly continuous functions, it is jointly continuous as well. Moreover, we can bound for $0 < \beta' < \beta$
\[ \left| \eu^{\beta' \, \norm{x-x'}} \, F(T)(x,x') \right| \le \sum_{n \geq 1} |a_n| \left| \eu^{\beta' \, \norm{x-x'}} \, T^n(x,x') \right| \le \sum_{n \geq 1} |a_n| \, C^n \]
uniformly in $x,x' \in \R^2$. Thus, the kernel of $F(T)$ is also exponentially localized.
\end{proof}

Considering \periodic\ operators, we immediately get the

\begin{corollary}
The set $\cS\sub{mp}^\infty$ of \periodic\ operators with JCEL integral kernels is a subset of $\mathcal{B}(L^2(\R^2))$ closed under adjoints, linear combinations, products, and analytic functional calculus via functions as in Proposition~\ref{prop exp loc int ker}\ref{item:Tiii}.
\end{corollary}

We show now that the operation of taking the commutator with a position operator defines a derivation on this $*$-subalgebra. 

\begin{proposition} \label{estimates for commutators} 
Let $X_j$ denote the $j$-th position operator, $j \in \set{1,2}$; let also $T \in \cS\sub{mp}^\infty$. Then the operator \([X_j, T]\), for $j \in \set{1,2}$, is in $\cS\sub{mp}^\infty$ as well. More generally, nested commutators of $T$ with a finite number of position operators
\[ [X_{j_1}, [\cdots , [X_{j_n}, T]] \cdots ]\,, \quad j_1,\ldots, j_n \in \set{1,2}\,, \]
are in $\cS\sub{mp}^\infty$ as well.
\end{proposition}
\begin{proof}
As for the integral kernel, it suffices to observe that
\[ [X_j, T](x,x') = (x_j-x_j') \, T(x,x')\,, \quad j \in \set{1,2}\,, \; x,x' \in \R^2\,, \]
is still jointly continuous and exponentially localized. Furthermore, commutativity with magnetic translation operators follows from the Jacobi identity
\[ \big[ [X_j, T], T_\gamma^A \big] = - \big[ [T, T_\gamma^A], X_j \big] - \big[ [T_\gamma^A, X_j], T \big] = \big[ [X_j , T_\gamma^A], T \big]\,, \quad \gamma \in \Gamma\,, \]
and the explicit expression for the commutator
\[ [X_j , T_\gamma^A] = \gamma_j \, T_\gamma^A \]
which can be immediately checked (cf.\ \cite[Lemma~2.1]{marcelli2021new}).

The same arguments can be iterated and applied to nested commutators.
\end{proof}

\section{Off-diagonal operators and Liouvillian super-operator} \label{sec:OD}

We start this Appendix by recalling the definition and properties of the diagonal and off-diagonal part of an operator in some Hilbert space $\mathcal{H}$ which is decomposed by an orthogonal projection $\Pi$ into the direct sum $\mathcal{H}= \Ran \Pi \oplus \Ran \Pi^\perp$, where $\Pi^\perp := \Id - \Pi$.

\begin{definition} \label{def:DOD}
Given an Hilbert space $\mathcal{H}$ and an orthogonal projection $\Pi$ on $\mathcal{H}$, define, for $A \in \mathcal{L}(\mathcal{H})$, 
\begin{itemize}
    \item the \emph{diagonal part} of $A$ as $A\su{D} := \Pi \, A \, \Pi + \Pi^\perp \, A \, \Pi^\perp$;
    \item the \emph{off-diagonal part} of $A$ as $A\su{OD}:= \Pi \, A \, \Pi^\perp + \Pi^\perp \, A \, \Pi$.
\end{itemize} 
We say that $A$ is \emph{diagonal} (respectively \emph{off-diagonal}) if $A = A\su{D}$ (respectively $A=A\su{OD}$).
\end{definition}

All statements in the next Proposition are immediate checks: compare also \cite[Sec.~6.1]{marcelli2021new}.

\begin{proposition} \label{4.1} 
The following hold.
\begin{enumerate}[label={(\roman*)},ref={(\roman*)}]
    \item An operator $A$ is diagonal if and only if it commutes with $\Pi$ (and therefore with $\Pi^\perp$). An operator $A$ is off-diagonal if and only if it satisfies $\Pi\, A = A\, \Pi^\perp$.
    \item \label{item:ABOD} Given operators $A$ and $B$, 
    \[ A = A\su{OD} \; \Longrightarrow \; (AB)\su{D} = A B\su{OD} \quad \text{and} \quad A = A\su{D} \; \Longrightarrow \; (AB)\su{OD} = A B\su{OD}\,. \]
    \item Each operator of the form $[\Pi, B]$ is off-diagonal.
    \item \label{item:AOD} For each operator $A$ it holds that
    \[A\su{OD}=\big[\Pi,[\Pi, A]\big]\,.\]
\end{enumerate}   
\end{proposition}

Operators in $L^2(\R^2)$ which are off-diagonal with respect to the Fermi projection $\Pi_0$ play a crucial role in the solution of the Liouville equation
\begin{equation} \label{liovillian eq}
    \mathcal{L}_{H_0}(B) := - \iu\, [H_0, B] = A
\end{equation}
as detailed in the next result.

\begin{proposition}\label{invlio}
Let $H_0$ be the Bloch--Landau Hamiltonian defined in~\eqref{eqn:BLH} and $\Pi_0$ be the corresponding Fermi projection~\eqref{eqn:Fermi}. Let $A \in \cS\sub{mp}^{\infty}$ be a \periodic\ operator on $L^2(\R^2)$ admitting a JCEL integral kernel, or $A=X_j$, $j \in \set{1,2}$, be a position operator. Then the following hold. 

\begin{enumerate}[label={(\roman*)},ref={(\roman*)}]
\item \label{item:ILi} The unique off-diagonal solution to the equation \eqref{liovillian eq} is given by 
\begin{equation} \label{eqn:L-1}
B = B\su{OD} = \mathcal{L}^{-1}_{H_0}(A) := \frac{1}{2\pi} \oint_\gamma \di z \, (H_0 - z)^{-1}\, [\Pi_0,A] \, (H_0 - z)^{-1}\
\end{equation}
where $\gamma$ is the same contour appearing in the Riesz formula~\eqref{eqn:Fermi}.
\item \label{item:ILii} It holds that $\mathcal{L}^{-1}_{H_0}(A\su{OD})=\mathcal{L}^{-1}_{H_0}(A)\su{OD}$.
\item \label{item:ILiii} The operator $\mathcal{L}^{-1}_{H_0}(A\su{OD})$ is in $\cS\sub{mp}^{\infty}$.
\end{enumerate}
\end{proposition}
\begin{proof}
For a proof of \ref{item:ILi}, we refer to~\cite[Lemma~4.1]{monaco2021stvreda} (compare also~\cite[Sec.~6.2]{marcelli2021new},~\cite[Appendix B]{marcelli2022purely} and references therein). 

Statement~\ref{item:ILii} easily follows from Proposition~\ref{4.1}.

It remains to prove \ref{item:ILiii}. For brevity, let us denote $T := A\su{OD}$: it is an operator in $\cS\sub{mp}^{\infty}$ in view of Propositions~\ref{prop exp loc int ker} and~\ref{estimates for commutators}. Let $(C_T,\beta_T)$ denote its localization constants, and $(C,\beta)$ be as in \eqref{eqn:L2CT}. Finally, let $0<\beta'< \min\{\beta, \beta_T\}$. For $x, x' \in \R^2$, we can then compute
\begin{align*}
2 \pi \, \Big| \eu^{\beta' \, \norm{x-x'}} &\, \mathcal{L}^{-1}_{H_0}(T)(x,x') \Big|  \le \sup_{z \in \gamma} \int \di x'' \, \di x''' \, \left| \eu^{\beta' \, \norm{x-x''}} \, R_{H_0}(z;x,x'') \right| \,  \left| \eu^{\beta' \, \norm{x''-x'''}} \, T(x'',x''') \right| \\
& \phantom{\mathcal{L}^{-1}_{H_0}(T)(x,x') \Big|  \le \sup_{z \in \gamma} \int \di x'' \, \di x''' \,} \left| \eu^{\beta' \, \norm{x'''-x'}} \, R_{H_0}(z;x''',x') \right| \\
& \le \sup_{z \in \gamma} \, \sup_{x,x' \in \R^2} \left| \eu^{\beta' \, \norm{x''-x'''}} \, T(x'',x''') \right| \, \left( \int \di x'' \, \left| \eu^{\beta' \, \norm{x-x''}} \, R_{H_0}(z;x,x'') \right| \right) \\
& \qquad\qquad \left(  \int \di x''' \, \left| \eu^{\beta' \, \norm{x'''-x'}} \, R_{H_0}(z;x''',x') \right| \right) \\
& \le C_T \sup_{z \in \gamma} \left(\sup_{x \in \R^2} \norm{\eu^{\beta \, \norm{x-\cdot}} \, R_{H_0}(z;x,\cdot)}_{L^2(\R^2)} \right) \\
& \qquad\qquad \left(\sup_{x' \in \R^2} \norm{\eu^{\beta \, \norm{\cdot - x'\cdot}} \, R_{H_0}(z;\cdot, x')}_{L^2(\R^2)} \right) \, \int \di y \, \eu^{-2\,(\beta-\beta') \,\norm{y}} \\
& \le C_T \, C^2 \, \int \di y \, \eu^{-2\,(\beta-\beta') \,\norm{y}}
\end{align*}
which is finite uniformly in $x, x' \in \R^2$. Therefore, the operator $\mathcal{L}^{-1}_{H_0}(T)$ admits an exponentially localized integral kernel.

To prove joint continuity, we argue as in~\cite[Appendix~A.1]{Moscolari_thesis}. Let $\set{\varphi_n}_{n \in \N}$ be an orthonormal basis of $L^2(\R^2)$. The integral kernel of $T$ can be expanded, as a function in $L^2(\R^2 \times \R^2)$, as
\[ T(x'',x''') = \sum_{n,m \in \N} c_{n,m} \, \varphi_n(x'') \, \varphi_m(x''')\,, \quad c_{n,m} \in \C\,.\]
Define, for $N \in \N$,
\[ T_N(x'',x''') := \sum_{0 \le n,m \le N} c_{n,m} \, \varphi_n(x'') \, \varphi_m(x''')\,. \]
Consequently
\begin{align*}
\sup_{x,x' \in \R^2} \left| \mathcal{L}^{-1}_{H_0}(T)(x,x') - \mathcal{L}^{-1}_{H_0}(T_N)(x,x') \right| & \le \norm{T - T_N}_{L^2(\R^2 \times \R^2)} \, \sup_{x \in \R^2} \norm{R_{H_0}(x,\cdot)}_{L^2(\R^2)} \\
& \qquad\qquad  \sup_{x' \in \R^2} \norm{R_{H_0}(\cdot,x')}_{L^2(\R^2)} \\
& \le C^2 \, \norm{T - T_N}_{L^2(\R^2 \times \R^2)}
\end{align*}
which goes to zero as $N \to \infty$. On the other hand, $\mathcal{L}^{-1}_{H_0}(T_N)(x,x')$ is a jointly continuous function of $x,x' \in \R^2$: indeed,
\begin{align*} 
\mathcal{L}^{-1}_{H_0}(T_N)(x,x') & = \sum_{0\le n,m\le N} c_{n,m} \, \oint_\gamma \frac{\di z}{2\pi} \, \left( \int \di x'' \, R_{H_0}(z;x,x'') \, \varphi_n(x'') \right)  \\
& \qquad \qquad \left( \int \di x''' \, R_{H_0}(z;x''',x') \, \varphi_m(x''') \right) \\
& = \sum_{0\le n,m\le N} c_{n,m} \, \oint_\gamma \frac{\di z}{2\pi} \, \left[R_{H_0}(z) \, \varphi_n\right](x)\, \left[R_{H_0}(\overline{z}) \, \varphi_m\right](x') \\ 
\end{align*}
and $R_{H_0}(z)$ maps $L^2$-functions to $H^2_A(\R^2) \subset C^0(\R^2)$, by the Sobolev embedding. We conclude that $\mathcal{L}^{-1}_{H_0}(T)(\cdot,\cdot)$ is a uniform limit of jointly continuous functions, and as such it is itself jointly continuous, as claimed.
\end{proof}

\section{Trace per unit area} \label{sec:TPUA}

The concept of trace per unit area (or per unit volume, in higher dimensions) plays a crucial role in the analysis of extended quantum systems, particularly when measuring expectation values of extensive observables in spatially infinite configurations. While the usual operator trace becomes ill-defined in such contexts, averaging over a finite box and taking the thermodynamic limit allows for the extraction of meaningful quantities such as the charge current density. In this Appendix, we introduce a class of operators admitting a well-defined trace per unit area and present its key algebraic properties. 

We begin with a preliminary observation.

\begin{remark} \label{magn_cov} 
Consider $T \in \mathcal{L}\sub{mp}$, and assume it admits a measurable integral kernel. The condition of commutation with magnetic translations implies the following magnetic periodicity condition at the level of its kernel:
\begin{equation} \label{mpc}
\eu^{-\iu \gamma \cdot A(x)} \, T(x-\gamma,x') = T(x, x'+\gamma) \, \eu^{-\iu \gamma \cdot A(x' + \gamma)},  \quad \gamma \in \Gamma\,, \quad x,x' \in \R^2\,.
\end{equation}
In particular, if the kernel is jointly continuous, then its restriction to the diagonal $\{x=x'\}$ is $\Gamma$-periodic:
\begin{align*} 
T(x+\gamma, x+\gamma) &= \eu^{\iu \gamma \cdot A(x+\gamma)} \, \left[ T(x+\gamma,x+\gamma) \eu^{-\iu \gamma \cdot A(x + \gamma)} \right] \\
& = \eu^{\iu \gamma \cdot A(x+\gamma)} \, \left[ \eu^{-\iu \gamma \cdot A(x+\gamma)} T(x+\gamma-\gamma,x) \right] = T(x,x)\,, \quad \gamma \in \Gamma\,.
\end{align*}
\end{remark}

\begin{definition}[Trace per unit area]
Let $T \in \mathcal{L}\sub{mp} \cap \mathcal{B}(L^2(\R^2))$ admit a measurable integral kernel. Let 
\begin{equation} \label{eqn:Lambda1}
\Lambda_1 := \set{y_1 \, a_1 + y_2 \, a_2 \in \R^2 : y_j \in [-1/2,1/2], \: j \in \set{1,2}}
\end{equation}
be the unit cell of the periodicity lattice $\Gamma$, and
\[ \Lambda_L := \bigcup_{\substack{\gamma \in \Gamma \\ |\gamma| \le L}} \Lambda_1 + \gamma\,. \]
We define its \emph{trace per unit area} to be
\begin{equation} \label{traccia puv su lambda 1}
    {\mathcal{T}}(T) := \lim_{L \to \infty}\frac{1}{|\Lambda_L|} \int_{\Lambda_L} \di x \, T(x,x) = \frac{1}{|\Lambda_1|} \int_{\Lambda_1} \di x \ T(x,x)
\end{equation}
where the last equality follows from the $\Gamma$-periodicity of $T(\cdot,\cdot)$.

We denote by $\cS^1\sub{mp}$ the set of operators $T$ such that $\mathcal{T}(T) < \infty$.
\end{definition}

\begin{proposition}[Properties of trace per unit area] \label{prop classe traccia puv}
\begin{enumerate}[label={(\roman*)},ref={(\roman*)}]
    \item \label{item:Tlinear} If  $T_1, T_2 \in \cS\sub{mp}^1$ and $\alpha \in \mathbb{C}$ then 
    \[{\mathcal{T}}(T_1 + \alpha \, T_2 )= {\mathcal{T}}(T_1)+ \alpha \, {\mathcal{T}}(T_2)\,.\]
    \item \label{item:Tcyclic} The trace per unit area is \emph{cyclic}: if $T_1 \, T_2, T_2 \,T_1 \in \cS\sub{mp}^1$ then
    \[\mathcal{T}(T_1\,T_2)=\mathcal{T}(T_2\,T_1).\]
    \item \label{item:Tinf} If $T\in \cS\sub{mp}^\infty$ then
    $\mathcal T(T)<\infty$, i.e.\ $T \in \cS\sub{mp}^1$.
\end{enumerate}
\end{proposition}

\begin{proof}
\ref{item:Tlinear} is trivially implied by the linearity of the integral. 

Cyclicity of the trace per unit area, as in \ref{item:Tcyclic},  follows from the magnetic periodicity condition stated in Remark~\ref{magn_cov}. Indeed, using Fubini's theorem, we have
\begin{equation} \label{eqn:T1T2=T2T1}
\begin{aligned}
\mathcal T(T_1 \, T_2)
&=\frac{1}{|\Lambda_1|}\,\int_{\Lambda_1}\int_{\R^2} T_1(x,z)\,T_2(z,x) \, \di z \, \di x = \frac{1}{|\Lambda_1|}\,\int_{\R^2}\int_{\Lambda_1} T_2(z,x) \, T_1(x,z) \, \di x \, \di z \\
&=\frac{1}{|\Lambda_1|} \sum_{\gamma \in \Gamma}\int_{\Lambda_1 + \gamma}\int_{\Lambda_1} T_2(z,x) \, T_1(x,z) \di x \, \di z \\
& = \frac{1}{|\Lambda_1|} \sum_{\gamma \in \Gamma}\int_{\Lambda_1 }\int_{\Lambda_1} T_2(y+ \gamma,x) \, T_1(x,y + \gamma) \, \di x\, \di y 
\end{aligned}
\end{equation}
where the third equality splits the $z$-integral on translates of the $\Gamma$-periodicity cell $\Lambda_1$. Now notice that, from \eqref{mpc}, it holds that
\begin{align*}
T_2(y+ \gamma,x) \, T_1(x,y + \gamma) &= \eu^{-\iu \gamma \cdot A(y)} \left[ \eu^{\iu \gamma \cdot A(y)}\, T_2(y+ \gamma,x) \right] \left[ T_1(x,y+ \gamma) \, \eu^{-\iu \gamma \cdot A(y+\gamma)}  \right] \, \eu^{\iu \gamma \cdot A(y+\gamma)}\\
&=\eu^{-\iu \gamma \cdot A(y)} \left[ T_2(y,x-\gamma) \, \eu^{\iu \gamma \cdot A(x-\gamma)} \right] \left[ \eu^{-\iu \gamma \cdot A(x)} \, T_1(x-\gamma,y) \right] \, \eu^{\iu \gamma \cdot A(y+\gamma)} \\
&= \eu^{\iu \gamma \cdot \phi(x,y,x-\gamma,y+\gamma)} \, T_2(y,x-\gamma) \, T_1(x-\gamma,y)\,,
\end{align*}
where
\[ \phi(x,y,x-\gamma,y+\gamma) := A(x-\gamma) - A(x) + A(y+\gamma) - A(y)\,, \quad x,y \in \Lambda_1, \: \gamma \in \Gamma.\]
The expression $\phi$ is linear in the vector field $A = A\sub{lin} + A\sub{per}$, so we can compute it separately for the linear part and the periodic part of the magnetic vector potential. The latter clearly gives a vanishing contribution, due to $\Gamma$-periodicity of $A\sub{per}$. Taking into account the linear part reduces the phase to
\[ \phi(x,y,x-\gamma,y+\gamma) := A(x) - A(\gamma) - A(x) + A(y) +A(\gamma) - A(y) = 0\,. \]
Due to this non-trivial phase cancellation, we can resume the computation in \eqref{eqn:T1T2=T2T1} and conclude
\begin{align*}
\mathcal T(T_1 \, T_2) & =  \frac{1}{|\Lambda_1|} \sum_{\gamma \in \Gamma}\int_{\Lambda_1 }\int_{\Lambda_1} T_2(y,x- \gamma) \, T_1(x - \gamma,y) \, \di x\, \di y \\
&=\frac{1}{|\Lambda_1|} \sum_{\gamma \in \Gamma}\int_{\Lambda_1 }\int_{\Lambda_1-\gamma} T_2(y,z) \, T_1(z,y) \, \di z\, \di y \\
&=\frac{1}{|\Lambda_1|}\int_{\Lambda_1 }\int_{\R^2} T_2(y,z) \, T_1(z,y) \, \di z\, \di y \\
& = \frac{1}{|\Lambda_1|}\int_{\Lambda_1 } (T_2 \, T_1)(y,y) = \mathcal{T}(T_2 \, T_1)\,,
\end{align*}
as wanted.

Finally, for what concerns point \ref{item:Tinf}, it suffices to notice operators in $\cS^\infty\sub{mp}$ have a jointly continuous kernel, so that its restriction to the diagonal $x \mapsto T(x,x)$ is integrable on $\Lambda_1$.
\end{proof}

Finally, we prove a fact which is instrumental to the proofs in Section~\ref{sec:main}: compare also~\cite[Lemma 2.1]{cornean2021beyond} and ~\cite[Proposition 3.3]{cornean2025on}.

\begin{proposition} \label{prop:T(UTU-1)}
If $T \in \cS^\infty\sub{mp}$ and $U \in \mathcal{U}(L^2(\R^2))$ is \periodic\ and such that $U-\Id$ has a JCEL integral kernel, then
\[ \mathcal{T}(U^{-1} T U) = \mathcal{T}(T)\,. \]
\end{proposition}
\begin{proof}
Write
\[ U^{-1} T U = (U^* - \Id) T U + T U = (U^* - \Id) T (U-\Id) + (U^*-\Id)T + T (U-\Id) + T \]
so that $U^{-1} T U \in \cS^\infty\sub{mp} \subset \cS^1\sub{mp}$. Upon observing that
\[(U-\Id) (U^*-\Id) = - (U-\Id) - (U^*-\Id) \]
we reach the conclusion upon using cyclicity of the trace per unit area.
\end{proof}

\begin{remark}
In view of Theorem~\ref{thm:NEASS} and the above Proposition, the NEASS $\Pi_n^\eps$ and the Fermi projection $\Pi_0$ produce the same \emph{integrated density of states}, i.e.\ have the same trace per unit area:
\[ \mathcal{T}(\Pi_n^\eps) = \mathcal{T}(\Pi_0)\,. \]
The latter is shown in~\cite{cornean2021beyond, cornean2025on} to be a linear function of the strength $b$ of the constant part of the magnetic field; moreover, the coefficient of $b$ (that is, the derivative of $\mathcal{T}(\Pi_0)$ with respect to $b$) equals, up to a factor $(2\pi)^{-1}$, the \emph{Chern marker} of the projection $\Pi_0$, which coincides with the double-commutator formula for the Hall conductivity $\sigma\sub{Hall}$ from~\eqref{sigmahall}. The above equality shows that the same conclusions hold for the NEASS $\Pi_n^\eps$.
\end{remark}

\bibliographystyle{siam}
\bibliography{citations}

\end{document}